\documentclass[preprint,12pt]{aastex}
\usepackage{natbib,epsfig}
\usepackage{graphicx}
\usepackage{multirow}
\usepackage{amsmath}

\begin{document}

\title{What do we know, what do we think we know and what do we not know about Gamma-Ray Bursts}
\author{Ehud Nakar}
\affil{Raymond and Beverly Sackler School of Physics \& Astronomy,
Tel Aviv University, Tel Aviv 69978, Israel}

\begin{abstract}
Decades of improving data and extensive theoretical research have
led to a popular model of gamma-ray bursts. According to this model,
a catastrophic event in a stellar system results in the formation of
a compact central engine, which releases a fraction of a solar
rest-mass energy within seconds in the form of ultra-relativistic
jets. Dissipation of the jets energy leads first to prompt gamma-ray
emission and later to a long lasting afterglow. Here I summarize the
introduction that I gave to the debate ``where do we stand?" in the
conference ``The Shocking Universe" held in Venice. This is a very
brief summary of my view of the facts that we are (almost) certain
about, models that are popular but may need rethinking, and main
open questions.
\end{abstract}

\section{introduction}
Gamma-Ray Burst (GRB) observations during the 1990's lead to the
impression that a GRB starts complex and becomes simpler to model
with time. The popular view was that a progenitor stellar system
goes through some catastrophic event, whose nature is unknown,
leading to the formation of a compact central engine. Somehow this
central engine deposits a fraction of a percent of the gravitational
energy of the system in the form of kinetic energy of a minute
fraction of its mass, thereby launching a collimated
ultra-relativistic outflow. Internal shocks (or another form of
internal interaction) within the outflow convert some of the outflow
energy into accelerated electrons and magnetic fields, which produce
the prompt emission that we observe, most likely via synchrotron
radiation. Later, an interaction of the outflow with the surrounding
medium drives a self-similar decelerating blast wave that
accelerates electrons and amplifies the ambient magnetic field to
produce the afterglow. The afterglow was considered relatively
simple to model and understand, since all the complex initial
conditions are washed out and only the initial energy and geometry
of the outflow determine the blast wave evolution.

Observations during the last decade have led on one hand to a
remarkable breakthrough in the identification of the progenitor
system, and on the other hand to the realization that  even the
afterglow physics is much more complex than suspected. Theoretical
modeling of the different physical processes improved and became
more detailed and accurate, but also more complicated. Additionally,
new ingredients were invoked to explain various specific
observations of specific bursts, significantly increasing the
model's degrees of freedom, sometimes to the point where the models
do not provide useful predictions anymore.

The purpose of the open discussion that I led in the conference was
to take a step back and evaluate where do we stand in the grand
picture. I
presented to the audience three questions:\\
\begin{itemize}
\item  What do you know with a very high confidence (almost as
facts) about the physics of GRBs? What observations are the base for your confidence? \\
\item Which of the models (for progenitors, prompt emission, afterglow,
etc.), if any, you think that are probably correct?\\
\item About which elements of GRB physics you think that we have almost no clue?
What do we need in order to learn about these?\\
\end{itemize}

Unfortunately no detailed notes were taken during the discussion so
here I will summarize only the introduction that I gave, which
includes my short answers to these questions. I divide my discussion
into five different stages of the explosion: progenitor, central
engine, outflow properties, prompt emission and afterglow. For each,
I give (a very partial) answer to the three questions. I touch only
the physics of the explosions themselves without considering the
study of the large scale environment (e.g., host galaxies). I
discuss long and short GRBs separately only in the ``progenitor"
section. The reason is that as far as we know the rest of the stages
may be (and possibly are) rather similar. Finally, one of the
comments during the discussion (by Chris Fryer) was that we are much
better at ruling out models than at proposing viable ones.
Nevertheless, I discuss here only models which may describe the
physics properly, without mentioning those that were ruled out.

Given the very broad scope of the discussion and the limited space
of this proceeding, I regretfully cannot give the appropriate credit
to the large number of observational and theoretical works on which
my views are based. Note that some of the ideas that are discussed
here were also presented by Maxim Lyutikov in his presentation,
which he summarizes in this proceeding.

\section{Progenitor}

\noindent {\it What do we confidently know?} \\
First, we know that there are at least two types of progenitors,
those of (the majority of) long GRBs and those of (the majority of)
short GRBs. This is based on three major observed differences
between the two GRB types. The host galaxies are very different. The
intrinsic redshift distribution is very different. An association
with a supernova (SN) was established for almost any nearby long
GRB, while a SN association was ruled out to any nearby short GRB.

We know much more about the progenitors of long GRBs. First, the
progenitor system includes a very massive star. This knowledge is
based on the association of specific nearby GRBs with SNe, on the
high specific star formation rate of the host galaxies and on the
location of the bursts within the most star forming regions of their
hosts. Second, {\it at least some} progenitors produce broad line Ic
SN (or SN like emission) within about $\pm 1$ day of the GRB.  The
evidence for this is not as straightforward as commonly assumed, and
although the evidence are quite convincing (see below), future may
hold surprises. Most of the nearby GRB-SNe events are not
necessarily genuine GRBs (and maybe not even genuine SNe, in the
sense that the driver of the explosion may be different than that of
typical core-collapse SNe). For example GRBs 980425 (SN 1998bw),
031203 (SN 2003lw), 060218 (SN 2006aj) and the very recent GRB
100316D (SN2010bh) all show a single pulse of soft gamma-rays which
contains about $10^{48}$ erg. These low luminosity bursts are very
different than typical cosmological GRBs and their $\gamma$-ray
emission is almost certainly produced by different physical
mechanism. GRB 030329/SN 2003dh was the first example where a SN was
directly associated with a GRB that is more similar to cosmological
ones. At z=0.16, GRB 030329 released an isotropic equivalent energy
of $\sim 10^{52}$ erg in two pulses of $\gamma$-rays. Until recently
this event was the only direct link between cosmological GRBs and
Ib/c SNe. Recently, Massimo Della Valle reported in an Astronomical
Telegram (No. 1602) of the detection of SN 2008hw associated with
GRB 081007 at z=0.53 (isotropic equivalent energy of $\sim 10^{51}$
erg in $\gamma$-rays), fortifying this connection. Apart for these
direct SN/GRB associations, there are also about a dozen cases where
a ``red bump", which is presumably contributed by the SN light, is
evident in the late time afterglow of cosmological GRBs, further
supporting the connection.

We do not know much about short GRB progenitors. All we know at a
high level of confidence is that they are different than long GRBs
and that at least some of the progenitor systems do not include
massive stars.\newline

\noindent {\it Popular models that need confirmation}\\
The most natural picture that explains why long GRB progenitors are
associated with massive stars and why they explode within $\pm 1$
day of a SN is that the GRB and the SN are produced simultaneously
following the collapse of a massive stellar core. While this is
probably true, there is no direct evidence that this is indeed the
case. Moreover, SNe modelers typically ignore the simultaneous
launch of GRB relativistic jets, while GRB central engine modelers
typically ignore the production of a SN. There is currently no model
that incorporates in detail the basic ingredients of both and the
interplay between them.

The most popular model for short GRB progenitors is neutron
start-neutron start or neutron start-black hole merger. But, while
being a very attractive model (especially for those interested in
gravitational wave signals), which gets a passing grade in the
comparison with all currently available observations, there is still
no conclusive (or even strong) evidence that this is indeed the
correct progenitor model.

\noindent {\it Some of the main open questions and how can we answer
them}\\
Three important open questions about long GRB progenitors are: What
is the role (if any) of various progenitor system properties
(metalicity, binarity, mass, etc.)? Are all long GRBs associated
with Ib/c SN? and is there a continuous transition from regular SNe
through low-luminosity GRB/SNe to cosmological GRBs? The answer to
all these questions needs mostly an accumulation of more
observations, where I want to stress the importance of searching for
SNe signal in an intermediate redshift ($z \sim 0.5-1$) GRBs. In
additional to improved observations, the key to third question is
theoretical modeling of the link between GRBs and SNe.

The main question about short GRB progenitors is simply what are
they. Obviously, the ultimate compact binary merger model test will
be via the detection or non-detection of  gravitational waves from
nearby short GRBs. But before we get to this stage, there is still
much that can be done. Here the focus should be to constrain the
environment and redshift distribution of the bursts. For this we
need a controlled large sample of bursts with redshift, host type
and location within the host. Additionally, deep limits on dark
X-ray afterglows can constrain the circum-burst density. Finally, we
are in a great need of a reliable classification scheme that can
tell the difference between (physically) short and long GRBs.

\section{Central engine}
\noindent {\it What do we confidently know?} \\
There is not much that we know with high confidence about the
central engine.  We basically know that it is a compact ($<10^7$ cm)
object that converts a fraction of the system's gravitational energy
into collimated relativistic jets continuously over a duration that
is much longer than its dynamical time.

\noindent {\it Popular models that need confirmation}\\
There are two popular models, with no conclusive evidence that
points to one of them (or that rules them out). The first, much more
popular model, is an accretion near the neutrino Eddington limit on
a stellar black hole. The main advantages of this model are that
similar accretion (although at much lower rates) is the known engine
of active galactic nuclei (AGNs) and $\mu$-quasars, and that it can,
in principle, release $10^{53}$ erg in the form of a relativistic
outflow. The main shortcoming of the model is that there are many
unknowns on the way that this engine works and there is no clear
idea why the specific properties of GRB outflows are generated.
Second, especially in short bursts, it is hard to explain late
engine activity.

The second model is a milli-sec proto-magnetar. Here, in the typical
version of the model, the gravitational energy is first converted
into rotational energy of the newly born magnetar and then released
in the form of a relativistic outflow. The main advantages of this
model are that, once the engine is formed, the physics of the
outflow launching is better understood and that it can explain late
engine activity rather naturally. A severe disadvantage of this
model is that it has a limited energy of $\approx 5 \cdot 10^{52}$
erg.

\noindent {\it Some of the main open questions and how can we answer
them}\\
As one can understand from the discussion above, there are many open
questions, including the most basic ones, such as what is the
physical outline of the engine. However proceeding in the quest to
understand the engine is very difficult, mostly because it is
completely hidden from us today. The reason is that the observed
electromagnetic radiation is produce above or near the photosphere
which is larger than the engine by many orders of magnitude. An
observable test that can be carried out today, as pointed out in the
discussion by Dale Frail, is the search for ultra-energetic bursts
($\sim 10^{53}$ erg), which if found will be hard to explain by the
proto-magnetar model.

A breakthrough in the understanding of the engine is expected if we
will be able to probe the engine directly, e.g., via gravitational
waves. Without such ground breaking observations it seems that the
major effort should be invested in theoretical and numerical
modeling, in order to better understand the models and to come up
with testable predictions. A breakthrough in the understanding of
AGN and/or $\mu$-quasar engines may help here as well.

\section{Outflow properties}
\noindent {\it What do we confidently know?} \\
We know that the outflow is ultra-relativistic. The main evidence
for relativistic motion comes from the requirement for low
$\gamma\gamma\rightarrow e^-e^+$ optical depth. In most bursts,
where photons above $\sim$MeV are not observed (most likely due to
insensitivity of the detectors), the outflow Lorentz factor,
$\Gamma$, must be larger than about $30$. Although extrapolation of
the observed spectrum to high ($\gg$MeV) frequencies suggest that
the true opacity limit is a few hundreds. There are also about half
a dozen bursts where observations of GeV photons set the lower limit
at about $1000$ (note that opacity limits may vary if the high
energy emission is not produced at the same place as the low energy
emission, e.g. prompt vs afterglow). There is no robust upper limit
to the outflow Lorentz factor (although afterglow theory suggests
that it cannot be much larger than $\approx 1000$). The most robust
confirmation that the outflow is indeed relativistic (although the
lower limit is $\Gamma
> 5$), is the measurement of the size of the radio afterglow of GRB
030329. The rather low lower limit is expected since the measurement
takes place at a time that, according to afterglow theory, the
outflow was significantly decelerated.

We are also quite certain that at least some long GRB outflows are
narrowly beamed (we do not know much about the collimation of short
GRBs). The list of arguments for collimation are composed of several
independent, strong, yet not conclusive, evidence. These are energy
requirements (without beaming some GRBs would release more than
$10^{55}$ erg), afterglow jet breaks and radio calorimetry.

Based on the constraints on the beaming we can deduce that the total
energy (corrected for beaming) carried by the outflow of long GRBs
is about $10^{50}-10^{52}$ erg and that the luminosity is $\sim
10^{49}-10^{51}$ erg/s. In the few cases where radio calorimetry can
be done, an energy $10^{51}-10^{52}$ erg is measured. In the case of
short GRBs we know only that the {\it isotropic equivalent} energy
emitted in $\gamma$-rays is about $10^{49}-10^{52}$ erg and the {\it
isotropic equivalent luminosity} is about $10^{50}-10^{52}$ erg/s.

\noindent {\it Some of the main open questions and how can we answer
them}\\
What is the actual Lorentz factor of the outflow? We have only
robust lower limits but the true value of the Lorentz factor is
unknown yet. It may be found if $\gamma\gamma$ attenuation will be
identified in the spectra, hopefully  by Fermi.

What is the detailed outflow geometry? Is it a top hat or does the
energy falls gradually with the angle from the axis? Is it patchy or
not? These questions are typically attacked using afterglow
observations, since during the prompt emission we observe only a
tiny patch of the outflow. I think that the progress here will come
from the observational side. The effort to resolve these questions
by modeling of afterglow jet-breaks requires many more simultaneous
radio-optical-X-ray afterglow observations while radio calorimetry
will hopefully improve significantly by the EVLA. Additional
observational tools that may be available in the future are a
statistically large sample of orphan afterglows and detailed
polarization measurements.

A long standing open question, which is crucial for the
understanding of both the central engine and the prompt emission, is
what component of the outflow is dominant energetically. Is it
baryonic, Poynting-flux or maybe pairs?  A prediction of the
baryonic outflow model was that Swift will observe many bursts with
bright optical flashes during the early afterglow. Swift detected
only a few optical flashes, thereby supporting a non-Baryonic
outflow. But, as it is often the case, non-detection does not
provide conclusive evidence. Currently, I do not have an idea of
future observations that may bring us closer to identify the outflow
composition. Part of the reason is that only the poorly understood
prompt emission and early afterglow are likely to be affected by the
outflow composition. Additionally, no specific model of the process
that converts Poynting-flux into the observed emission is available
for comparison to observations. Thus, a theoretical progress on this
front will be helpful.

\section{Prompt emission}
\noindent {\it What do we confidently know?} \\
We know almost nothing for certain. The most robust statements,
which may prove to be wrong in the future, are as follows. Based on
opacity arguments, the emission originates at radius $> 10^{11-12}$
cm while interaction with the external medium dictates that it takes
place at a radius $<10^{16-17}$ cm. Energy requirements and
afterglow modeling implies that  the prompt emission is very
efficient, $\gtrsim 10\%$, in converting the outflow energy into
sub-MeV $\gamma$-rays. Finally, high variability points strongly
towards a dissipation mechanism that is internal to the flow,
although there are suggested models of pointing flux dominated
outflows, where the internal dissipation is triggered by interaction
with the external medium.

\noindent {\it Popular models that need confirmation}\\
The most popular model is the internal shock model, where the
outflow dissipation is done by hydrodynamical shocks between
different portions of the flow. The main advantage of the model is
that it naturally explains the high variability. But the
disadvantages of the model made it less popular in the last several
years. The main one is the limited efficiency. Despite of a large
theoretical effort, there is no consensus on a natural way to bypass
this problem. Additionally, there are many properties of the prompt
emission that are not explained naturally in the internal shock
model.

\noindent {\it Some of the main open questions and how can we answer
them}\\
Despite of the impressive set of prompt emission observations, the
entire topic is one big question mark. With thousands of superb
light curves and spectra we cannot confidently identify even the
dominant radiation process, not to speak on nailing down the
dissipation process or clearly find out the origin of a large number
of unexplained detailed properties of the emission (e.g., energy
dependent pulse shapes, $E_p$ distribution, various correlations
etc.).

The main reason for the difficulty is that the high energy power-law
spectrum suggests that the emission is generated above the
photosphere, where the main candidate is synchrotron radiation by
ultra-relativistic electrons. However, the steep low energy spectrum
is very difficult to explain with synchrotron and it suggests a
thermal component at the base, which is modified above or near the
photosphere by inverse Compton. However the smooth transition from
the low to high energy spectrum implies that for this model to work
the electrons must be at most mildly relativistic and that they
should carry a comparable energy to the radiation. This suggests
that a very efficient dissipation of the outflow energy takes place
just below the photosphere. Otherwise, if the energy is dissipated
too deep it is lost to adiabatic expansion, while if it is
dissipated high above the photosphere the thermal component (if it
exists) does not interact strongly with the electrons. So far there
is no consensus on a mechanism that naturally provides these
requirements (although lately some interesting new candidates where
suggested), and that can naturally explain a large fraction of the
observed features.

Observationally we already have great $10$ keV - $1$ MeV data set
and a large number of optical - X-ray observations that coincide
with the end of the prompt emission of long GRBs. The only hope that
I can see for observational breakthrough in the near future is by
Fermi, which already provided useful information that strongly
disfavor inverse Compton by ultra-relativistic electrons as the
source of the prompt emission. But, so far, Fermi observations did
not pointed towards the correct model, while they did raise up new
open questions. If such breakthrough will not take place, then a
theoretical study of the existing models that are still viable and
development of new ones, are the most promising way to understand
the prompt emission.

\section{Afterglow}
\noindent {\it What do we confidently know?} \\
The late afterglow is generated during (and almost certainly by) the
interaction  with the circum burst medium. This statement is
robustly based on the decelerated expansion of the afterglow image
of GRB 030329. It is also strongly supported by the light curve and
spectral afterglow evolution, which show a continuous power-law
decay (in both the flux and peak frequency) and variability time
scales that increase with time.

\noindent {\it Popular models that need confirmation?}\\
By far, the most popular afterglow model is the external forward
shock model, and in my view it is very likely that the afterglow
emission, at least starting a few hours after the burst, is
generated by forward external shock. The major success of this model
is that with a very simple parametrization  (of only a few free
parameters) it encompasses the gross observed afterglow properties
over eight orders of magnitude in frequency and four orders of
magnitude in time. However, an examination of the fine details shows
that the model is far from being complete. The simple model is not
compatible with the detailed observations of many bursts (e.g., the
exact spectral and temporal power-law indices). There is a set of
observations (especially during the first $10{^3}-10^{4}$ s), which
are difficult to explain by forward shock emission even by
significantly complicating the simple model. Some examples are X-ray
flares, X-ray plateaus and some chromatic breaks. Different
extensions of the basic model are invoked to explain the observed
deviation from the model. However, often these are tailored on a
burst to burst basis, and most dangerously, in some cases modelers
provide their models with so much freedom so they lose their
predictability and with it their usefulness. Finally, the simple
model simply parameterizes the unknown microphysics, which may have
a very complicated behavior, by three constants that has (almost) no
theoretical first principle predictions.

\noindent {\it Some of the main open questions and how can we answer them}\\
Is the external shock model correct? I expect that the strongest
evidence on this point will come from unique events, such as GRB
030329. Until these are observed, more detailed multi-wavelength
(radio - X-ray) observations will be helpful. There are many optical
(and some radio) afterglows without detailed X-ray coverage before
the Swift era. Now, that Swift-XRT provides impressive X-ray
light-curves, there are too few optical light curves and almost no
radio afterglow detections.

What is the cause of the X-ray plateau and chromatic breaks? These
are among the most surprising Swift observations, which still waits
for a theoretical breakthrough, whose direction I cannot predict.

Plasma microphysics is among the toughest topics of any
astrophysical environment and GRBs are not different. However, if
the afterglow is generated by an external shock, then this is a
relatively clean system where the initial conditions are rather
simple: an ultra-relativistic collisionless shock that propagates
into a very weakly magnetized electron-proton plasma. Understanding
the microphysical processes that take place in such system will
requires an extensive numerical work accompanied by careful
theoretical analysis. This field made an impressive progress in
recent years and I hope that first principle theory of GRB
microphysics will be developed during the next decade. Detailed GeV
observations may also prove useful to farther constrain the
microphysical properties of the radiating plasma.

\section{Conclusions}
A quick scan of the GRB "facts" listed here shows that among the
observations on which we base these facts, the role of single unique
events (e.g., GRB 030329) is similar to, or maybe even larger than,
that of the large burst samples. It also shows that we are still
ignorant even about fundamental ingredients of the explosions. Even
afterglow modeling, which before the launch of Swift was thought by
many to be satisfactory, is now being reconsidered. Nevertheless, if
I compare today's knowledge to what we knew a decade ago the
progress is impressive. I hope that the coming decade will be as
eventful as the passing one and that ten years from now we will be
able to say that our understanding has improved by at least as much.

I am in debt of all the conference participants which contributed to
the debate. I am grateful to Guido Chincarini for the worm and
welcoming hospitality during the conference and to Tsvi Piran and
Carles Badenes for useful comments on the manuscript.

\end{document}